\documentclass[letter]{aa} 
\usepackage{txfonts}
\usepackage{graphicx, graphics}
\usepackage{CJK}
\usepackage{bbm}
\usepackage{amsmath, amssymb, bm}
\usepackage[nodayofweek]{datetime}
\input{prepcitationformat.txt}
\usepackage{txfonts}
\usepackage{changes}

\usepackage{mathtools}

\DeclarePairedDelimiterX\braket[2]{\langle}{\rangle}{#1 \delimsize\vert #2}

\usepackage[switch]{lineno} 

\usepackage{tikz}

\definecolor{lime}{HTML}{A6CE39}
\DeclareRobustCommand{\orcidicon}{%
    \raisebox{-3pt}{\begin{tikzpicture}
    \filldraw [lime, yshift=-2pt] (0, 0) circle [radius=0.16]
    node[white] {\raisebox{1pt}{\hspace{0.5pt}\fontfamily{qag}\selectfont\tiny i\scalebox{0.8}{D}}};
    \end{tikzpicture}}
    \hspace{-2.5mm}
    \vspace{-0.25pt}
}

\global\def\tablenotemark#1{{\color{blue}{\normalfont\textsuperscript{\scriptsize #1}}}} 

\newdateformat{monthyearday}{
  \THEYEAR\ \monthname[\THEMONTH] \twodigit{\THEDAY}}
\newdateformat{daymonthyear}{%
  \twodigit{\THEDAY}\ \monthname[\THEMONTH] \THEYEAR}

\newcommand{\orcidauthor}[2]{#2\href{http://orcid.org/#1}{\orcidicon}}
\DeclareRobustCommand{\ion}[2]{%
\relax\ifmmode
\ifx\testbx\f@series
{\mathbf{#1\,\mathsc{#2}}}\else
{\mathrm{#1\,\mathsc{#2}}}\fi
\else\textup{#1\,{\mdseries\textsc{#2}}}%
\fi}

\titlerunning{3C~273 with \textit{HST}/STIS Coronagraphy}
\authorrunning{Ren et al.}

\begin{document}
\begin{CJK*}{UTF8}{gbsn}
\title{3C~273 Host Galaxy with \textit{Hubble Space Telescope} Coronagraphy\thanks{FITS images for Figs.~\ref{fig-3C273}--\ref{fig-3C273-isophote} are only available at the CDS via anonymous ftp to \url{cdsarc.cds.unistra.fr} (\url{130.79.128.5}) or via \url{https://cdsarc.cds.unistra.fr/viz-bin/cat/J/A+A/}}}

\author{
\orcidauthor{0000-0003-1698-9696}{Bin B. Ren (任彬)}\thanks{Marie Sk\l odowska-Curie Fellow}\inst{\ref{inst-oca}, \ref{inst-uga}, \ref{inst-cit}, \ref{inst-jhu}, \ref{inst-jhu-ams}}
\and
\orcidauthor{0000-0002-2691-2476}{Kevin Fogarty}\inst{\ref{inst-ames}, \ref{inst-cit}}
\and
\orcidauthor{0000-0002-1783-8817}{John H. Debes}\inst{\ref{inst-stsci}}
\and
\orcidauthor{0000-0002-7676-9962}{Eileen T. Meyer}\inst{\ref{inst-umbc}}
\and
Youbin Mo\inst{\ref{inst-ucsd}}\thanks{Now at Google.}
\and
\orcidauthor{0000-0002-8895-4735}{Dimitri Mawet}\inst{\ref{inst-cit}, \ref{inst-jpl}}
\and
\orcidauthor{0000-0002-3191-8151}{Marshall D. Perrin}\inst{\ref{inst-stsci}}
\and
Patrick M. Ogle\inst{\ref{inst-stsci}}
\and
\orcidauthor{0000-0001-9525-3673}{Johannes Sahlmann}\inst{\ref{inst-rhea}}
}

\institute{
Universit\'{e} C\^{o}te d'Azur, Observatoire de la C\^{o}te d'Azur, CNRS, Laboratoire Lagrange, Bd de l'Observatoire, CS 34229, F-06304 Nice cedex 4, France; 
\url{bin.ren@oca.eu} \label{inst-oca}
\and
Universit\'{e} Grenoble Alpes, CNRS, Institut de Plan\'{e}tologie et d'Astrophysique (IPAG), F-38000 Grenoble, France \label{inst-uga}
\and
Department of Astronomy, California Institute of Technology, MC 249-17, 1200 E California Blvd, Pasadena, CA 91125, USA \label{inst-cit}
\and
Department of Physics and Astronomy, The Johns Hopkins University, 3701 San Martin Drive, Baltimore, MD 21218, USA \label{inst-jhu}
\and
Department of Applied Mathematics and Statistics, The Johns Hopkins University, 3400 N Charles St, Baltimore, MD 21218, USA \label{inst-jhu-ams}
\and
NASA Ames Research Center, Moffett Field, CA 94035, USA; \url{kevin.w.fogarty@nasa.gov} \label{inst-ames}
\and
Space Telescope Science Institute (STScI), 3700 San Martin Drive, Baltimore, MD 21218, USA \label{inst-stsci}
\and
Department of Physics, University of Maryland, Baltimore County, 1000 Hilltop Circle, Baltimore, MD 21250, USA \label{inst-umbc}
\and
Department of Physics, University of California, San Diego, CA 92093, USA \label{inst-ucsd}
\and
Jet Propulsion Laboratory, California Institute of Technology, 4800 Oak Grove Drive, Pasadena, CA 91109, USA \label{inst-jpl}
\and
RHEA Group for the European Space Agency (ESA), European Space Astronomy Centre (ESAC), Camino Bajo del Castillo s/n, E-28692 Villanueva de la Ca\~nada, Madrid, Spain \label{inst-rhea}
}

\date{Received 12 October 2023 / Revised 14 February 2024 / Accepted 14 February 2024}

\abstract
{The close-in regions of bright quasars' host galaxies have been difficult to image due to the overwhelming light from the quasars. With coronagraphic observations in visible light using the Space Telescope Imaging Spectrograph (STIS) on the \textit{Hubble Space Telescope}, we removed 3C~273 quasar light using color-matching reference stars. The observations revealed the host galaxy from $60\arcsec$ to $0\farcs2$ with nearly full angular coverage. Isophote modeling revealed a new core jet, a core blob, and multiple smaller-scale blobs within $2\farcs5$. The blobs could potentially be satellite galaxies or infalling materials towards the central quasar. Using archival STIS data, we constrained the apparent motion of its large scale jets over a $22$~yr timeline. By resolving the 3C~273 host galaxy with STIS, our study validates the coronagraph usage on extragalactic sources in obtaining new insights into the central ${\sim}$kpc regions of quasar hosts.}


\maketitle

\section{Introduction}
Quasars are unique laboratories for the extreme physics governing active galactic nuclei (AGN) accretion and feedback, and are important drivers for galaxy evolution and enrichment \citep[e.g.,][]{sijacki07, mcnamara12, moustakas19}. However, since the central source in a quasar can have a visual luminosity comparable to the entire host galaxy within which it resides, the point spread function (PSF) of a quasar's central source -- when seen with a telescope with finite mirror size -- often dominates the light at inner ${\sim}$kpc scales. Many features of quasars' circumnuclear regions \citep[e.g.,][]{almeida17}, such as inflows, dusty tori, winds, and jets, have visual and infrared components that are overwhelmed as a result \citep[e.g.][]{ford94, ford14}.  Moreover, the dynamics and morphology of the host galaxy close-in to the nuclear region are likewise ``swamped out'' in visible and near-infrared (NIR) observations. While advances in radio interferometry have allowed us to glimpse the event horizons of two supermassive black holes at ${\sim}0.1$~mas scale and  their surrounding environments (i.e., M87: \citealp{eht19}, Sagittarius A*: \citealp{eht22, Lu2023}), many processes critical to feeding and feedback in extremely luminous quasars will only be understood once we obtain high-contrast, high-resolution imaging in the visible and NIR \citep[e.g.,][]{martel03, gratadour15, metis,moustakas19, Rouan19, Grosset21, Ding2023}. 

To study quasar hosts, infrared and radio interferometry are capable of imaging the inner several parsecs of the circumnuclear region, and can therefore look at the broad line region, inner radius of the torus, and jets on this scale \citep[e.g.][]{kishimoto11, lister13, gravity18}. However, circumnuclear disk, infalling material, and jet activity in the narrow-line region are best observed at an intermediate scale in the further out regions. To study these regions, observations of naturally dust-obscured quasars (``natural coronagraphs''; e.g., \citealp{jaffe93, vandermarel98}) only preferentially sample these structures in disk-like hosts or during epochs of peak dust production early-on in mergers, giving an incomplete picture of quasar evolution \citep{urrutia08, schawinski12, moro18}. The Space Telescope Imaging Spectrograph (STIS) coronagraph onboard the \textit{Hubble Space Telescope} (\textit{HST}), however, can fill the gap at ${>}0.5$~kpc to ${\sim}$kpc scales and larger to study the morphology of host galaxies. Visual imaging of the sub-kpc structures of the dust, jet, and host galaxies of quasars with STIS coronagraph provides a unique opportunity to complement ground- and space-based infrared high-contrast imaging with Keck and \textit{JWST}. 

The prototypical quasar 3C~273 was first identified based on its redshift of $0.158$ by \citet{schmidt63}. With the central quasar dominating the signals across visible to radio wavelengths and thus overwhelming the host galaxy, the study of the latter makes it necessary to first properly remove the quasar light. In visible wavelengths, \citet{martel03} placed the 3C~273 behind a coronagraph using \textit{HST}/ACS, removed the quasar light using reference star images, and revealed the host galaxy in visible light exterior to an angular radius of ${\sim}1\farcs5$. With non-coronagraphic imaging after empirical PSF removal, these non-asymmetric signals persist after host galaxy modeling \citep[][Figure~10 therein]{zhang19}. In (sub)-millimeter observations \citet{komugi22} subtracted point source models  using ALMA, and revealed the surroundings where the millimeter continuum emission colocates with the extended emission line region in [\ion{O}{III}] observed with VLT/MUSE in \citet{husemann19}. These efforts unveiled complex structure for the 3C~273 host from multiple aspects, calling for dedicated imaging that would better reveal and characterize the 3C~273 host with available state-of-the-art instruments. Here we report our coronagraphic high-contrast imaging observations using \textit{HST}/STIS, where we reached an inner working angle (IWA) of ${\sim}0\farcs2$ in visible light to reveal the 3C~273 host galaxy.

\section{Observation and Data Reduction}\label{sec-obs}

The \textit{HST}/STIS coronagraph offers broadband imaging in visible through NIR light ($0.2~\mu$m -- $1.15~\mu$m; \citealp{stisihb}). Its narrowest occulter BAR5 can image the surroundings of bright sources exterior to ${\sim}0\farcs2$ from the center  \citep{schneider17, debes19}. To reveal the surroundings of bright central sources using STIS,  a careful selection of reference stars is needed to avoid color mismatch given its broadband \citep{debes19}, otherwise a non-matching PSF can induce spurious signals \citep[][Figure~8 therein]{ren17}. A reference star should be of similar color, magnitude, and in close proximity of a science target to maximize the success in coronagraphic imaging \citep{debes19}. To explore the inner regions of the 3C~273 host galaxy down to ${\sim}0\farcs2$, or ${\sim}0.5$~kpc,\footnote{$0\farcs1=0.27$~kpc for 3C~273 in $\Lambda$CDM cosmology with $\Omega_m=0.27$, $\Omega_\Lambda=0.73$, and $H_0=71$km~s$^{-1}$~Mpc$^{-1}$.} we observed it with two reference stars using \textit{HST}/STIS in \textit{HST} GO-16715 (PI: B.~Ren). The two reference stars serve as the empirical PSF templates to remove the 3C~273 quasar light.


\subsection{PSF reference star selection}
On the lower Earth orbit, \textit{HST} is affected by instrumental effects such as breathing caused by changes in the Solar angle, radiation, and Earth's shadow. These effects on STIS observations could be empirically captured and removed when well-chosen PSF stars are close in angle to the science target. To effectively remove the PSF from the central source, \citet{debes19} recommended close match in magnitude and color in the $B$ and $V$ band. With the high-sensitivity space-based measurements in visible light from the \textit{Gaia} mission \citep[DR3:][]{gaiadr3}, \citet{walker21} showed that magnitude and color match in \textit{Gaia} filters may also provide reasonably good PSFs for STIS observations. 

Using \textit{Gaia}~DR3, we selected two reference stars -- TYC~287-284-1 (hereafter ``PSF1'') and TYC~292-743-1 (hereafter ``PSF2'') -- to serve as the PSF templates for 3C~273 (hereafter ``science target''), see Appendix~\ref{app-coron}.  For 3C~273, its \textit{Gaia}~DR3 magnitude $G = 12.84$, color $Bp - Rp = 0.494$ and $G - Rp = 0.348$. Within $4\fdg7$ from 3C~273, the chosen PSF1 has \textit{Gaia}~DR3 magnitude $G = 12.77$, color $Bp - Rp = 0.515$ and $G - Rp = 0.327$. Within $2\fdg7$ from 3C~273, the chosen PSF2 has magnitude $G = 11.488$, color $Bp - Rp = 0.536$ and $G - Rp = 0.340$. By choosing two PSF reference stars with such faintness that do not have existing infrared excess measurements, with infrared excess being indicative of circumstellar disks \citep[e.g.,][]{cotton16}, we can reduce the probability that a reference star might host a circumstellar disk that negatively impacts the PSF removal for 3C~273. In addition, this strategy reduces the possibility that background objects, which are beyond the detection limits of exiting instruments, negatively impact the PSF removal of 3C~273.

\begin{figure*}[htb!]
	\includegraphics[width=\textwidth]{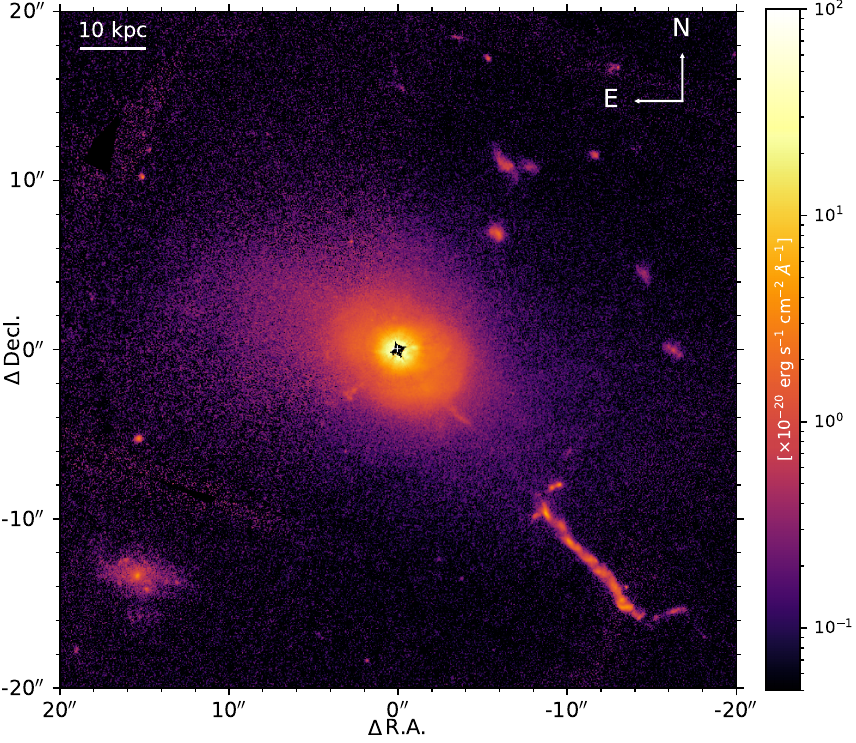}
    \caption{3C~273 host galaxy and surroundings in visible light seen with the \textit{HST}/STIS coronagraph. The surface brightness is in log scale.}
    \label{fig-3C273}
    
    (The data used to create this figure are available.)
\end{figure*}

\subsection{Observation}
Coronagraphic imaging with \textit{HST}/STIS relies on the blockage of light in central regions using its physical occulters. In STIS, BAR5 and WedgeA0.6 are two nearly perpendicular occulting locations that offer IWAs of ${\sim}0\farcs2$ and ${\sim}0\farcs3$, respectively \citep{stisihb}. To enable a full angular coverage of extended structures, however, the two locations are near to the edges of the field of view of STIS, making it unrealistic to roll the telescope ${\sim}90^\circ$ to achieve a nearly $360^\circ$ coverage given the scheduling limits of \textit{HST} using four consecutive orbits (e.g., Figure~3 of \citealp{debes19}),\footnote{For observation planning, see Phase 2 of Astronomer's Proposal Tool (APT) for actual roll ranges for given observation times.} see Appendix~\ref{app-coron}. Therefore, we scheduled two sets of observations in GO-16715, and each set is composed of 4 contiguous ``back-to-back'' \textit{HST} orbits, see Table~\ref{tab:sys-log} for the observation log. By observing 3C~273 at two different epochs spanning ${\sim}2$ months in \textit{HST} Cycle 29, the relative roll is $84\fdg032$ between the central visits of the two epochs to approach a full angular coverage.

In GO-16715, we observed only one object in an orbital visit, with a sequence of ``target-target-PSF-target'' in a 4-orbit observation set. This ensures that the telescope thermal distribution is stabilized when a PSF star is observed. The observations were in CCD Gain = 4 mode to permit high dynamic range imaging \citep[e.g.,][]{debes19}. For the 3 orbits on the target, we rolled the telescope by either ${\pm}15^\circ$ (UT 2022-01-08) or ${\pm}5^\circ$ (UT 2022-03-26)\footnote{Roll angle of ${\pm}5^\circ$ due to updated scheduling constraints with the \textit{HST} guide star catalog then. When permitted, ${\pm}22\fdg5$ rolls can maximize angular coverage by avoiding \textit{HST} diffraction spikes that are along the (off)-diagonal directions in STIS images.\label{footnote-obs}} to approach a ${\sim}360^\circ$ angular coverage for it, see Fig.~\ref{fig-coverage} for the coverage map of 3C~273.

Within one science target orbit, we observed 3C~273 using both the BAR5 and the WedgeA0.6 occulting location with 3 readouts each. With three readouts each being a $315$~s exposure at an occulting location, the STIS flat-fielded files can identify cosmic rays for random noise removal. There are a total of 36 readouts from 6 orbits. For the reference stars, on the one hand, the \textit{Gaia}~DR3 $G$ magnitude of PSF1 is similar as that of the science target, with PSF1 being $0.07$ mag brighter. The observation strategy of PSF1 is identical to the target: to reach similar detector counts, each readout of PSF1 is $294$~s. There are 6 readouts in total for PSF1. On the other hand, the \textit{Gaia}~DR3 $G$ magnitude of PSF2 is $1.352$ mag brighter than that of the science target. To reach similar detector counts as 3C~273, each readout of PSF2 is $125$~s. Given the available time in one \textit{HST} orbit, the relatively shorter readout time permit the dithering of the telescope: the on-sky step is $0.25$ STIS pixel to reduce the impact from the non-repeatability of \textit{HST} pointing on STIS results \citep[e.g.,][]{ren17, debes19}. At each occulting location, we dithered the telescope twice with each dithering location having two $125$~s readouts, totaling 6 readouts per occulting location. There are a total of 12 readouts for PSF2.

\subsection{Data reduction}
\subsubsection{Pre-processing}
In long readouts ($315$~s for the target), the observation data quality with STIS may be compromised due to different noise sources (e.g., cosmic ray, shot noise, charge transfer inefficiency: CTI). To address this in data reduction, we first used the \texttt{stis\_cti} package\footnote{\url{https://pythonhosted.org/stis_cti/}} that applies the \citet{anderson10} correction to remove CTI effects for STIS CCD.\footnote{See \url{https://github.com/spacetelescope/STIS-Notebooks} for a usage example of \texttt{stis\_cti} under DrizzlePac.} We then used the data quality map in the CTI-corrected flat-fielded files, and followed \citet{ren17} to perform a median bad pixel replacement for the data that have been marked as a bad pixel (e.g., pixels with dark rate more than $5\sigma$ the median dark level, bad pixel in reference file, and pixels identified in cosmic ray rejection) around its $3{\times}3$-pixel neighbors. To correct for the geometric distortion in 2-dimensional STIS CCD images \citep{stiscal}, we then rectified the CTI-corrected images using the \texttt{x2d} function from \texttt{stistools}.\footnote{\url{https://stistools.readthedocs.io/en/latest/x2d.html}}

In STIS coronagraphic imaging, the central star is blocked by the STIS occulters, we thus used the two diffraction spikes to align the images \citep[e.g.,][]{schneider14, ren17}. We used the ``X marks the spot'' method \citep[e.g.,][]{schneider14}: we first fit Gaussian profiles to each column or row of the observation to identify the peak of the diffraction spikes, then fit lines and obtain the intersection point as the location of the star. In this way, we can obtain both the locations and their associated uncertainties from the observed data.

\subsubsection{Post-processing: PSF removal}\label{sec-pp}

For each target image, we minimized the standard deviation for the regions containing diffraction spikes after PSF subtraction to obtain a residual image. Specifically, we used the algorithmic mask from \citet{debes17} to identify the regions that are blocked by the STIS occulters in the observations. At each occulting location, we scaled the median of all the PSF readouts, and subtracted it from the target readouts to inspect the residual images. We derotated each residual map to north-up and east-left using its corresponding {\tt ORIENTAT} header, and used the element-wise median of all 36 residual maps (18 from each occulting location) as the final image for 3C~273 host galaxy. We multiplied the final image in units of counts~s$^{-1}$~pixel$^{-1}$ by  \texttt{PHOTFLAM} = $4.22\times10^{-19}$, which is the inverse sensitivity parameter recorded in the FITS file headers, to obtain the calibrated final image in units of erg~s$^{-1}$~cm$^{-2}$~$\AA^{-1}$, see Fig.~\ref{fig-3C273}. We also calculated the standard deviation for 
the regions that do not host signals in the final image to obtain the signal-to-noise map, see Fig.~\ref{fig-sn}.

To inspect the PSF removal quality dependence on different PSF references, we have experimented using a single PSF star for all the images, and we did observe a decrease in signal-to-noise from combining the two PSF stars. In fact, in the observations for PSF1, we identified at least two faint background sources within the STIS field of view (e.g., Fig.~\ref{fig-reduction-demo-psf}). Nevertheless, this was anticipated in our observation planning, and we minimized such risks by using multiple PSF stars in Table~\ref{tab:sys-log}. 

\begin{figure*}[bht!]
	\centering
	\includegraphics[width=0.85\textwidth]{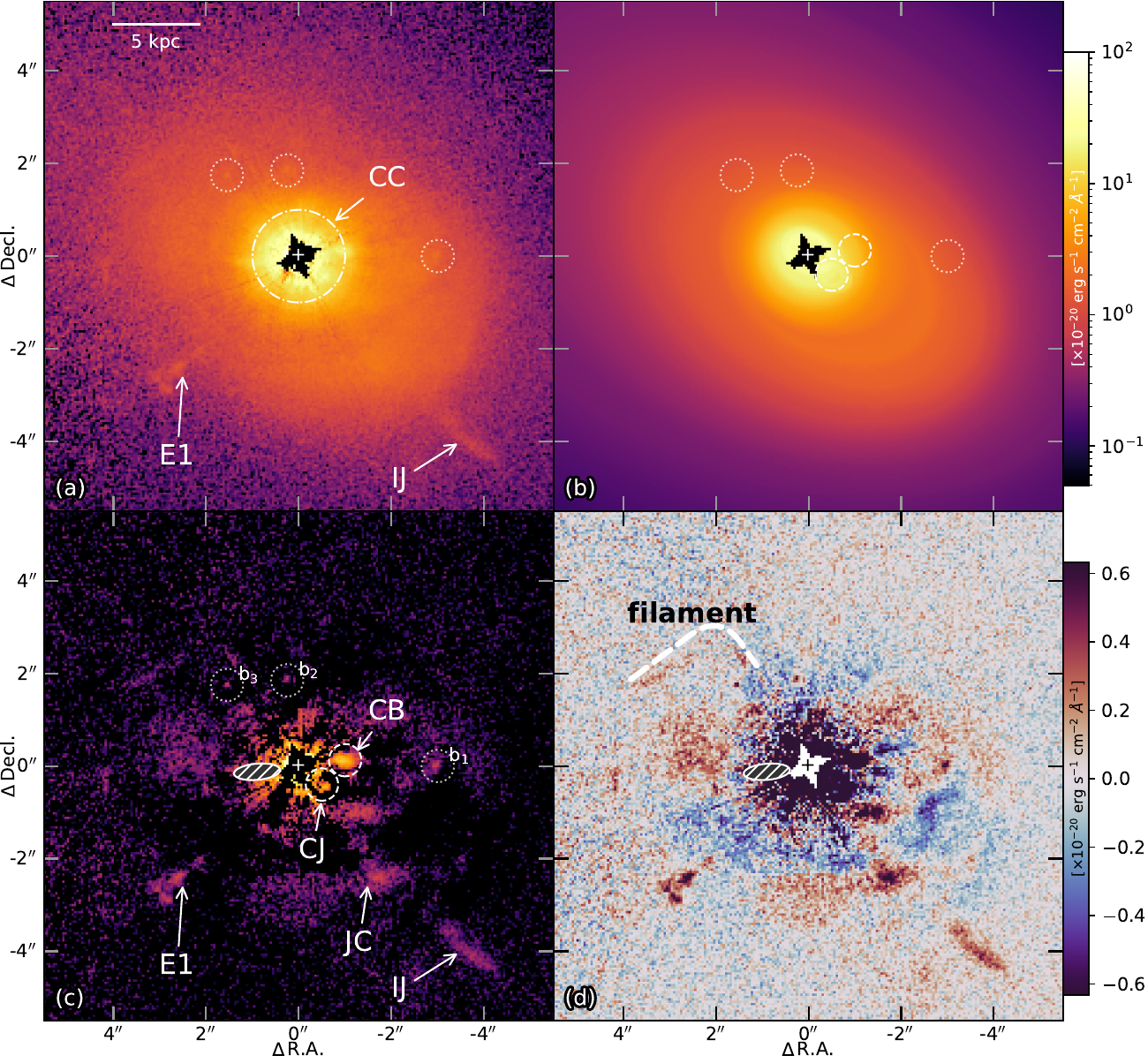}
    \caption{Host galaxy of 3C~273 within $5\arcsec$. (a) contains original data. (b) is the isophote model. (c) and (d) are isophote-removed data. We have newly identified a symmetric core component (CC) within ${\sim}1\arcsec$ (marked with dash dotted circle), a core blob (CB) component at ${\sim}1\arcsec$ to the west of the quasar, a core jet (CJ) component, three smaller scale blobs (marked with dotted circles) at ${\sim}4\sigma$ levels in comparison with their surrounding host galaxy, and filamentary structures at ${\sim}3\sigma$ levels. We recover the \citet{martel03} findings including jet component (JC), inner jet (IJ), and E1 component. Note: we masked out a PSF feature \citep[e.g.,][]{grady03} in (c) and (d) a shaded ellipse to the east of the center.}
    \label{fig-3C273-isophote}
        
\raggedright
(The data used to create this figure are available.)
\end{figure*}

\section{Results}\label{sec-res}

\subsection{3C~273 host}
\subsubsection{Isophote fitting}
We present in Fig.~\ref{fig-3C273-isophote} the host galaxy within $5\arcsec$ from the quasar. 
At the largest spatial scale in Fig.~\ref{fig-3C273}, we confirm the existence of two components -- the outer component (OC) and the inner component (IC) -- as previously identified in isophotic contours in \citet{martel03} with \textit{HST}/ACS. The OC is more extended on the north-east side than the IC in STIS wavelengths, with both centers offset from the central quasar, as has been reported in \citet{martel03}. 
With STIS coronagraphy in Fig.~\ref{fig-3C273-isophote}, we confirm the existence of the E1, jet component (JC), and inner jet (IJ) components identified in \citet{martel03}. 

To reveal small scale structures, we performed isophote modeling to enhance the visibility of structures of smaller spatial scale. We used the isophote package from \texttt{photuils} \citep{photutils}, which implemented the \citet{jedrzejewski87} method to iteratively fit elliptical isophotes to the 3C~273 host image. We present the isophote model in Fig.~\ref{fig-3C273-isophote}(b) and the model-removed residuals in Fig.~\ref{fig-3C273-isophote}(c), where we annotated the structures. To better reveal these structures, we reperformed isophote fitting while excluding them to reduce fitting bias, see the final isophote model and residuals in Fig.~\ref{fig-3C273-isophote}.

\subsubsection{New close-in host structures}
We identify a core blob (CB) component at ${\sim}1\arcsec$ to the west of the quasar in Fig.~\ref{fig-3C273-isophote}(c).  We also identify a core jet (CJ) component along the direction of the large scale jet. We additionally detect small-scale blobs spanning from $2\arcsec$ to $4\arcsec$ from the quasar. These blobs are marked with dotted circles in Fig.~\ref{fig-3C273-isophote}. We also detected a more symmetric core component (CC) for the host galaxy of 3C~273 within ${\approx}1''$ in Fig.~\ref{fig-3C273-isophote}(a). The surface brightness\footnote{The uncertainties in this Letter are $1\sigma$ unless otherwise specified.} is $(5\pm2)\times10^{-20}$~erg~s$^{-1}$~cm$^{-2}$~$\AA^{-1}$ for CB, $(14\pm5)\times10^{-20}$~erg~s$^{-1}$~cm$^{-2}$~$\AA^{-1}$ for CC, and $(3\pm2)\times10^{-20}$~erg~s$^{-1}$~cm$^{-2}$~$\AA^{-1}$ for CJ. In comparison, the background is $(1.3\pm1.3)\times10^{-22}$~erg~s$^{-1}$~cm$^{-2}$~$\AA^{-1}$, which is two orders of magnitude fainter than these new close-in structures. We present the measurements in Table~\ref{tab:sb}.

We mark the credible new structures in Fig.~\ref{fig-3C273-isophote}(c). First, the CB component is not a random or color-mismatch residual, in the sense that it does not have a radially shaped elongation that suggests color mismatch or telescope breathing \citep[e.g.,][]{schneider14}, and that its globular morphology does not resemble the spurious residuals in the existing STIS data archive for circumstellar structures in \citet{ren17}. Second, the CJ component is not only along the direction for the jet, but also of elliptical shape in the isophote model in Fig.~\ref{fig-3C273-isophote}(b). Third, the blobs (i.e., b$_1$, b$_2$, b$_3$) are not risen from random noise and they are ${\sim}4\sigma$ above their surrounding host galaxy (blob surface brightness: ${\sim}4\times10^{-21}$~erg~s$^{-1}$~cm$^{-2}$~$\AA^{-1}$): they persist in the individual reduction results from different telescope rolls in Fig.~\ref{fig-features-demo}, and they are not behind diffraction spikes regions where random noise may manifest positive residuals.
By fitting the surface brightness of CC with a bivariate normal distribution while ignoring these small-scale structures, we obtain a semi-major axis of $a_{\rm CC} = 0\farcs578\pm0\farcs005$ and a semi-minor axis of $b_{\rm CC} = 0\farcs477\pm0\farcs004$, and a position angle of $56\fdg2\pm1\fdg5$ for the major axis. The fitted center is located at $\Delta{\rm R.A.}=0\farcs031\pm0\farcs004$ and $\Delta{\rm Decl.}=-0\farcs024\pm0\farcs003$, or $0\farcs039\pm0\farcs004$ from the quasar, which is less offset from the quasar than IC and OC \citep[$0\farcs65$--$1\farcs40$;][]{martel03}. The ellipticity is $\eta_{\rm CC} = (a_{\rm CC} - b_{\rm CC})/a_{\rm CC} = 0.175\pm0.010$, which is smaller than the $\eta \gtrsim 0.3$ measurements in further out regions \citep[][Fig.~4 therein]{martel03}, suggesting that the host galaxy is more symmetric when it is closer to the quasar.

We detected evidence of extended filamentary structures to the Northeast, East and West of the galactic nucleus. These structures are visible in the residuals image (filament surface brightness: ${\sim}2\times10^{-21}$~erg~s$^{-1}$~cm$^{-2}$~$\AA^{-1}$) in Fig.~\ref{fig-3C273-isophote}(c)(d), obtained after computing and subtracting an elliptical isophote model of the underlying early-type galaxy. Candidate filaments extending ${\sim}5$--$10$ kpc to the Northeast are reminiscent of the emission line nebulae seen in e.g. the brightest cluster galaxy in the Phoenix cluster \citep{2018McDonald_CoolingFlowProblem} and believed to be multiphase gas that condenses out of the circumgalactic medium and fuels further AGN feedback \citep{2015Gaspari_AGNFeedback, 2018Gaspari_CCA}. Structures seen to the East several kpc from the galactic nucleus and several kpc to the West are also potentially consistent with this scenario, although either deeper or (ideally) spectroscopic follow-up observations will be necessary to characterize these structures. Further multi-band photometric or spectroscopic follow-up observations with \textit{JWST} will characterize these structures and determine the role they play in the lifecycle of AGN feedback in 3C 273.

\subsection{Jet motion}\label{sec-jet-motion}
3C~273 has been imaged with STIS, with the unocculted 50CCD imaging configuration, on 2000 April 3 in \textit{HST} GO-8233 (PI: S.~Baum). We do not remove the PSF for this observation using the coronagraphic archive from \citet{ren17}, since unocculted PSFs do not resemble the coronagraphic ones \citep{grady03}. Nevertheless, the jet is visible in both the 2006 and our 2022 observations, which establish a $7950$--$8027$~day separation, or $22$~year, for apparent motion measurement.

We aligned the rectified non-coronagraphic observations of 3C~273 with STIS using the ``X marks the spot'' method. In comparison with the coronagraphic observations, there are no data quality extensions in the flat-fielded files in \textit{HST} GO-8233, we thus remove the cosmic ray noises using the \texttt{Astro-SCRAPPY} code \citep{astroscrappy} that implements the \citet{vanDokkum01} approach in \texttt{astropy} \citep{astropy}. In comparison with \citet{meyer16} where two \textit{HST} instruments (WFPC2/PC and ACS/WFC) were used to obtain jet motion, our study with an identical instrument permits motion measurements with less offsets from different instruments.


To measure the jet motion between the two epochs using two images with different quality, we adopted the concept of dummy variables to simultaneously fit elliptical morphology and offset. Specifically, for one jet component, we fit identical bivariate normal distribution to its data in two epochs while allowing for translation and rotation, see Appendix~\ref{app-jm} for the details.

We also applied the same procedure to measure the offset for background sources to correct for motion biases due to different observations. The field rotation between the two epochs is $0\fdg088\pm0.004$, or $0\fdg0040\pm0\fdg0002$~yr$^{-1}$. This rotation rate is faster than the \citet{wardduong22} study of $0\fdg0031\pm0\fdg0001$~yr$^{-1}$, potentially due to our combination of coronagraphic images after rotation along the center of the quasar. However, our result confirms the trend of STIS long-term rotational evolution of the STIS CCD images.

\begin{figure}[htb!]
	\includegraphics[width=0.5\textwidth]{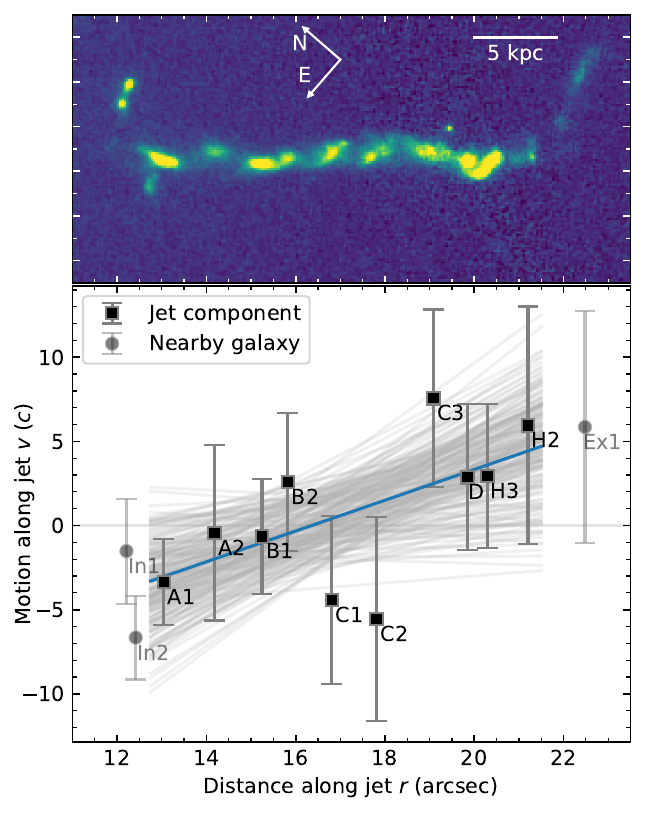}
    \caption{Motion of different components $v$ along the jet direction $r$ in units of speed of light $c$. The best-fit of $\frac{v}{c} = 0.9_{-0.5}^{+0.5}\left(\frac{r}{1\arcsec}\right) - 15_{-8}^{+8}$  in Sect.~\ref{sec-jet-motion} suggests faster outwards motion when the jet components are further from the quasar. The gray lines are 200 random samples from the best-fit parameters and their covariance.}
    \label{fig-motion}
\raggedright

\end{figure}

By adopting a conversion factor of $\frac{8.9856~c}{{\rm mas~yr}^{-1}}$ as in \citet{meyer16}, where $c$ is the speed of light, we calculated the apparent proper motion of the jet components along the jet axis, which is at a position angle of $-139\fdg0\pm1\fdg9$ east of north based on the jet components. We present the motion rates in Fig.~\ref{fig-motion} and Table~\ref{tab:motion} following the annotations in \citet{marshall01}. The motion of the jet components along the jet direction is consistent with zero at ${<}2\sigma$ levels. We witness a potential trend that the motion is faster when it is further out. By fitting a linear relationship between the distance to the quasar $r$ in arcsec and motion $v$ in $c$, and excluding the nearby galaxy components (i.e., In1, In2, and Ex1; e.g., \citealp{meyer16}), we have $\frac{v}{c} = 0.9_{-0.5}^{+0.5}\left(\frac{r}{1\arcsec}\right) - 15_{-8}^{+8}$.

\section{Summary}\label{sec-sum}
With the most recent understanding of the coronagraphic imaging capabilities of \textit{HST}/STIS in IWA and instrumentation stability \citep{debes19}, we have validated STIS coronagraphy using extragalactic observations in this study. By applying STIS coronagraph to the iconic quasar 3C~273, and capturing its coronagraphic PSF using two color-matching close-in reference stars, we have revealed its surrounding regions, including its host galaxy, from ${\sim}0\farcs2$ to ${\sim}60\arcsec$.

We have detected a more symmetric core component, CC, for the host galaxy of 3C~273, in addition to confirming the existing large-scale asymmetric components IC and OC that were previously identified in \textit{HST}/ACS coronagraphy from \citet{martel03}. With the STIS coronagraphic observations, we also identify a core blob (CB) component, as well as other point-source-like objects, after removing isophotes from the host galaxy. The nature of the newly identified components, as well as the point-source-like objects, would require observations from other telescopes for further study.

Using a $22$~yr timeline, we have constrained the apparent motion for the jet components of 3C~273. With an identical instrument between the two observations, we obtain that the jet components in the largest scale jet likely have faster motion when they are further from the quasar. We also confirm the long-term rotational trend of the STIS CCD images.

To characterize the observed host components for 3C~273, follow-up efforts in the near-infrared will help constrain the nature of them in multi-wavelength color and/or spectroscopy. From the ground, with the visible magnitude of quasars beyond the capability of most of coronagraphic imaging systems that perform adaptive optics correction in visible light, the usage of pyramid wavefront sensing in near-infrared wavelengths is necessary (Keck/NIRC2: \citealp{bond20}; VLT/ERIS: \citealp{davies18, kravchenko22}). From the space, \textit{JWST} coronagraphic and non-coronagraphic imaging efforts will also access such hosts, with the coronagraphic imaging observations necessary to probe deeper into the quasar surroundings.

Being the only operating space-based coronagraph in broadband visible light, \textit{HST}/STIS offers an IWA of ${\sim}0\farcs2$ to image the surrounding environment around bright central sources. In addition to carefully selected PSF stars for this study, with more reference star images taken at the smallest IWA positions after the commissioning of the BAR5 occulter for STIS \citep{schneider14}, the usage of a library of PSF image \citep{ren17} could better help in reducing the impact of color difference and telescope status. What is more, to reveal faint and extended structures that could reach the sensitivity limit of the STIS CCD, we recommend correcting its periodic readout variations \citep[temporal variation:][]{jansen03, jansen13} before image rectification. Moving forward, an ongoing calibration program, \textit{HST} GO-17135 
for STIS coronagraphy adopting a carefully designed dithering strategy might provide smaller IWAs, is expected to potentially push towards imaging the components that are behind the current coronagraphic occulters for stars and quasars towards ${\sim}0\farcs1$ (i.e., ${\sim}0.3$~kpc for 3C~273). Similarly, for \textit{JWST}, such attempts are also necessary given its supported sizes of the coronagraphs are significantly larger than $0\farcs1$ (e.g., \textit{JWST} GO-3087). With smaller IWAs for both telescopes, we can both confirm the existence of closest-in components and constrain their physical properties from multi-band imaging. In high-energy observations, we can better characterize such structures, as well as binary active galactic nuclei \citep[e.g.,][]{Pfeifle23}, when coronagraphic instruments are available in the future.

\begin{acknowledgements}
We thank the anonymous referee for constructive suggestions that improved this Letter. B.B.R.~would like to acknowledge the pioneering leadership of the late M.~Schmidt, whose office at the Cahill Center in Caltech was assigned to B.B.R.~in 2019, for inspiring future generations with his groundbreaking discovery of 3C~273. We thank Dean Hines, Anand Sivaramakrishnan, and Hsiang-Chih Hwang for discussions in the initial preparation of the observation proposal. We thank Paul Kalas, Tom Esposito, Yuguang Chen, and Zhi-Xiang Zhang for discussions. Based on observations with the NASA/ESA \textit{Hubble Space Telescope} obtained at the Space Telescope Science Institute, which is operated by the Association of Universities for Research in Astronomy, Inc., under NASA contract NAS5-26555. Support for program number \textit{HST} GO-16715 was provided through a grant from the STScI under NASA contract NAS5-26555. This project has received funding from the European Union's Horizon Europe research and innovation programme under the Marie Sk\l odowska-Curie grant agreement No.~101103114. This project has received funding from the European Research Council (ERC) under the European Union's Horizon 2020 research and innovation programme (PROTOPLANETS, grant agreement No.~101002188). This work has made use of data from the European Space Agency (ESA) mission {\it Gaia} (\url{https://www.cosmos.esa.int/gaia}), processed by the {\it Gaia} Data Processing and Analysis Consortium (DPAC, \url{https://www.cosmos.esa.int/web/gaia/dpac/consortium}). Funding for the DPAC has been provided by national institutions, in particular the institutions participating in the {\it Gaia} Multilateral Agreement. This research made use of Photutils, an Astropy package for
detection and photometry of astronomical sources \citep{photutils}.
\end{acknowledgements}

\bibliography{refs}

\appendix

\section{\textit{HST}/STIS coronagraphic imaging}\label{app-coron}
\begin{figure*}[htb!]
\centering
	\includegraphics[width=\textwidth]{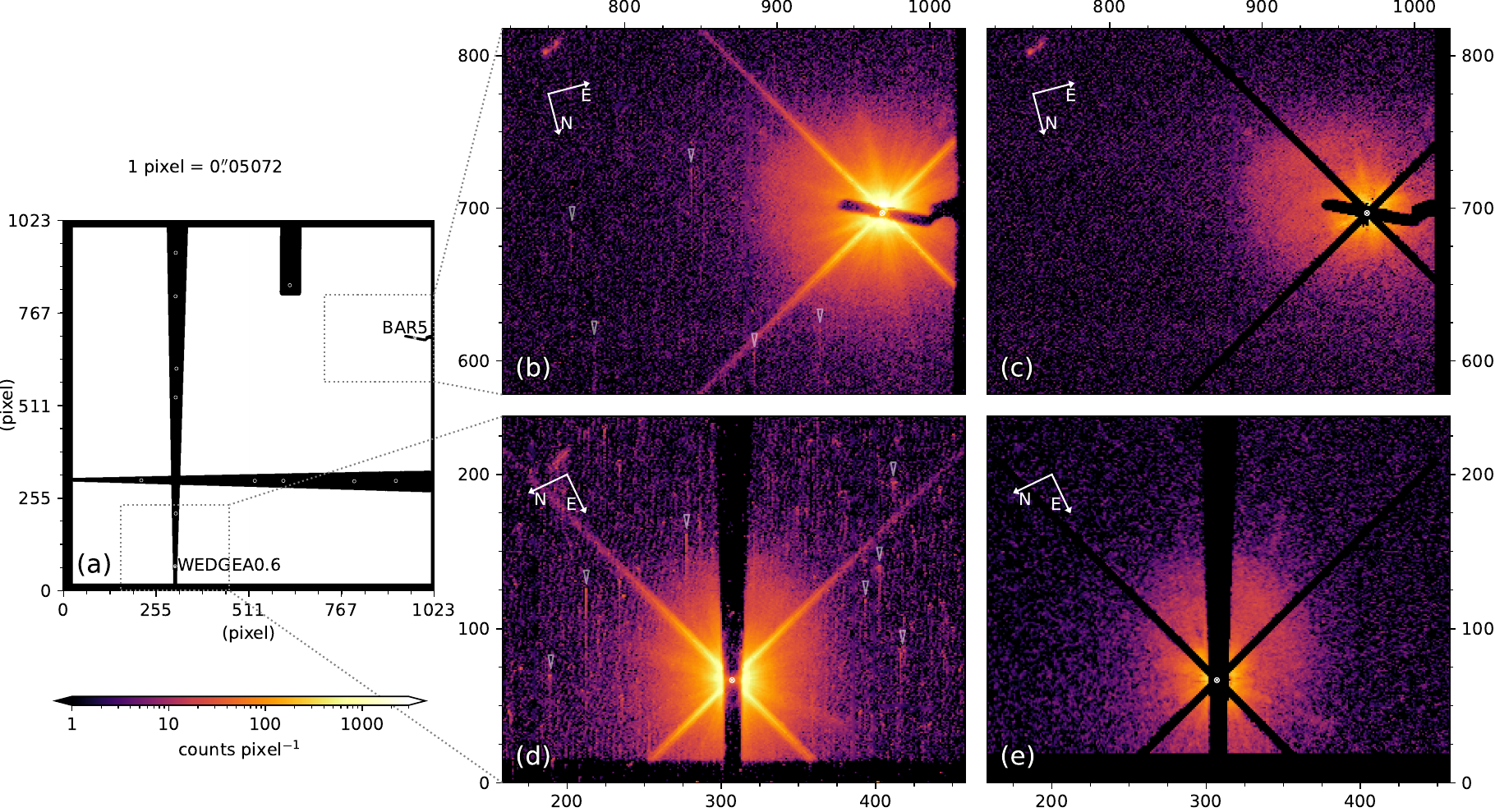}
    \caption{\textit{HST}/STIS coronagraphic imaging can reveal faint structures that are overwhelmed by the PSF of central sources. (a) STIS coronagraphic occulter from \citet{debes17}, with empirical occulting locations in \citet{ren17}. (b) and (d) are example 3C~273 exposures with a few CTI trail examples marked using triangles from the BAR5 and WEDGEA0.6 occulters, respectively. (c) and (e) are the PSF-removed results on rectified CTI-corrected images for (b) and (d).}
    \label{fig-hci-demo}
\end{figure*}

\textit{HST}/STIS coronagraphic imaging relies on well-chosen PSF templates for data reduction \citep[e.g.,][]{debes19}. We list the target and the PSFs in this study, along with their information and observation log, in Table~\ref{tab:sys-log}. We present the coronagraphic imaging occulting locations, as well as the coronagraphic data before and after PSF-removal, in Fig.~\ref{fig-hci-demo}.

The coronagraphic imaging mode of STIS offers four major occulters: the two perpendicular wedges (WedgeA and WedgeB), BAR10, and BAR5 which offers the smallest inner working angle of ${\sim}0\farcs2$ \citep[e.g.,][]{stisihb}. There are 13 default occulting locations, see Fig.~\ref{fig-hci-demo}(a) for their empirical centers from \citet{ren17}. In addition, STIS users can issue instrument offsets from these centers for customized imaging (i.e., ``POSTARG''). Among the 13 occulting locations, WedgeA1.0 (IWA ${\approx}0\farcs5$) and WedgeA0.6 (IWA ${\approx}0\farcs3$) have been mostly used in existing locations \citep{ren17}, thus offering the largest PSF template archives for PSF removal.

BAR5 is oriented ${\sim}12^\circ$ clockwise from a horizontal occulter, and we can image the surrounding environments of central sources down to ${\sim}0\farcs2$. In fact, due to occulter deformation pre-launch, the BAR5 occulter was not enabled for \textit{HST} General Observers science until \citet{schneider17}. In comparison, WedgeA is positioned vertically with WedgeA0.6 offering a ${\sim}0\farcs3$ IWA. With BAR5 additionally blocking nearly horizontal regions and WedgeA0.6 blocking vertical regions due to the existence of the coronagraphic occulters,  the combination of WedgeA0.6 and BAR5 is thus necessary to offer a nearly $360^\circ$ angular coverage (with an exception for the diagonal and off-diagonal diffraction spikes) with the smallest IWAs. In addition, on the one hand, WedgeA0.6 is located $68$~pixel or $3\farcs45$ from the bottom of the coronagraphic edge, and on the other hand, BAR5 is located $54$~pixel or $2\farcs74$ from the right coronagraphic edge, thus multiple telescope rolls can help image structures beyond these angular radii.

\begin{figure}[htb!] 
\centering
	\includegraphics[width=0.49\textwidth]{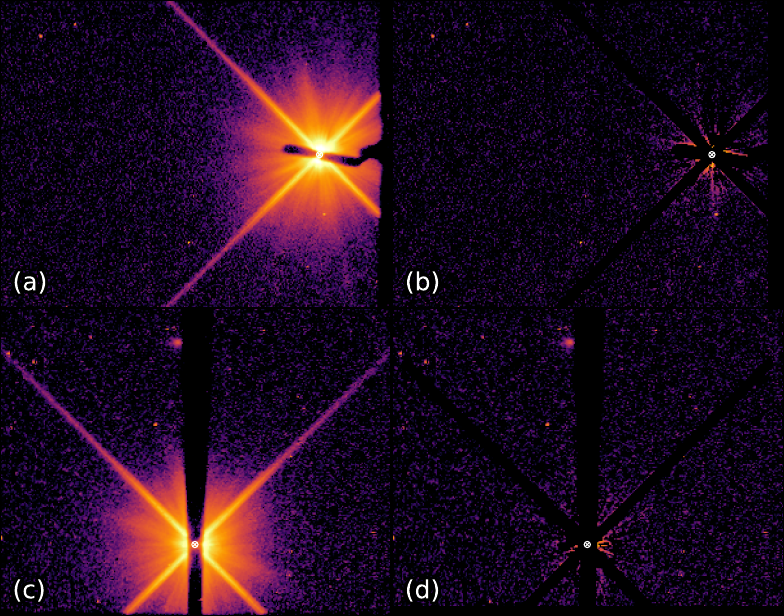}
    \caption{PSF removal for PSF1 using PSF2. (a) and (c) are rectified CTI-removed PSF1 exposures under BAR5 and WedgeA0.6, respectively. (b) and (d) are the corresponding PSF-removed residuals for PSF1 for (a) and (c). In comparison with the 3C~273 residuals in Fig.~\ref{fig-hci-demo} with identical color bars, there are no significant residuals for PSF1 that resemble 3C~273 residuals.}
    \label{fig-reduction-demo-psf}
\end{figure}

To demonstrate the improvement in PSF removal with coronagraphic imaging, we present in Fig.~\ref{fig-hci-demo}(b) and (d) two exposures on the detector frame at different roll angles using BAR5 and WedgeA0.6, respectively. With CTI effect corrected, Fig.~\ref{fig-hci-demo}(c) and (e) are the corresponding 3C~273 surroundings after PSF removal. The count rates can be reduced by a factor of ${\sim}5$ at ${\sim}1\farcs5$ after PSF removal, thus extracting the surroundings that are overwhelmed by the PSF. In addition, the count rate reduction is performed on the PSF halo of the central source behind the coronagraph, making another step forward in actual suppression of the central source light using a combination of instrumentation and data reduction. To validate the PSF removal results for 3C~273, we reduced the PSF1 exposures using the PSF2 exposures for a comparison by following an identical PSF removal process as 3C~273. In the PSF-removed residuals for PSF1 in Fig.~\ref{fig-reduction-demo-psf}, no significant signals resembling the 3C~273 residuals exist.

To obtain the final image in Fig.~\ref{fig-3C273}, multiple telescope rolls are necessary in obtaining a ${\sim}360^\circ$ angular coverage to resolve the constraints from the occulting locations being located near the edge of the STIS field of view and diffraction spikes. In addition, STIS allows subarray readouts in obtaining smaller field of view \citep[e.g.,][]{schneider18} to increase observation efficiency in reducing readout time in its electronics. We did not request the subarray readout here, since the purpose is to image the 3C~273 surroundings to the largest spatial extent.

\begin{table}[htb!]

\setlength{\tabcolsep}{1.75pt}
\caption{System parameters and STIS observation log\label{tab:sys-log}}
\hspace{-0.25cm}
\begin{tabular}{cc|ccc}    \hline\hline
\multicolumn{2}{c|}{} & {Target} & {PSF1} & {PSF2}\\ \hline
\multicolumn{2}{c|}{Object} & {3C~273} & {TYC~287-284-1} & {TYC~292-743-1} \\
\multicolumn{2}{c|}{$G$} & $12.84$ & $12.77$ & $11.49$ \\
\multicolumn{2}{c|}{$Bp-Rp$} & $0.494$ & $0.515$ & $0.536$ \\
\multicolumn{2}{c|}{$G-Rp$} & $0.348$ & $0.327$ & $0.340$ \\
\multicolumn{2}{c|}{Angular distance\tablenotemark{a}} & $\cdots$ & $4\fdg7$ & $2\fdg7$ \\ \hline
Visit & \texttt{ORIENTAT}\tablenotemark{b} & \multicolumn{3}{c}{Exposure time (UT 2022-03-26)} \\
01 & $165\fdg359$ & $2\times3\times315$~s & $\cdots$& $\cdots$ \\
02\tablenotemark{c} & $160\fdg359$ & $2\times3\times315$~s & $\cdots$& $\cdots$ \\
03 & $138\fdg984$ & $\cdots$ &  $2\times3\times294$~s & $\cdots$ \\
04 & $155\fdg359$ & $2\times3\times315$~s & $\cdots$& $\cdots$ \\ \cline{3-5} 
 &  & \multicolumn{3}{c}{Exposure time (UT 2022-01-08)} \\
05 & $-100\fdg609$ & $2\times3\times315$~s & $\cdots$& $\cdots$ \\
06\tablenotemark{c} & $-115\fdg609$ & $2\times3\times315$~s & $\cdots$& $\cdots$ \\
07 & $-112\fdg999$ & $\cdots$& $\cdots$ &  $2\times6\times125$~s  \\
08 & $-130\fdg609$ & $2\times3\times315$~s & $\cdots$& $\cdots$ \\ \hline
\multicolumn{2}{c|}{Total exposure} & $11,340$~s & $1764$s & $1500$~s\\ \hline
\end{tabular}
\begin{flushleft}
{\tiny \textbf{Notes}: 
$^a${\textit{Gaia}~DR3 angular distance to 3C~273.} $^b${The \texttt{ORIENTAT} parameter denotes the telescope roll angle using the angle the detector $y$ axis makes with North \citep{stisdhb}.} $^c${Central visits of the two observation sets. The telescope roll between the two visits is $84\fdg032$, which is used to approach a full angular coverage since the BAR5 and WedgeA0.6 occulting locations are close to the edges of the STIS detector.}
}
\end{flushleft}
\end{table}

In the post-processing of the STIS coronagraphic imaging data in Sect.~\ref{sec-pp}, we used the median exposures from two stars as the empirical PSF template for 3C~273. Alternatively, the usage of archival data may better capture the PSFs of central sources to extract extended structures \citep[e.g.,][]{soummer14, ren18, sanghi22, xie23}. To explore this, we updated the coronagraphic PSF archive at the BAR5 location for STIS from \citet{ren17}. With both the principal-component-analysis-based PSF modeling approach \citep[][]{soummer12} and the non-negative matrix factorization method \citep[][]{ren18} applied to STIS coronagraphic imaging (e.g., \citealp{ren17, ren18, walker21}), we did not observe significant improvement in data reduction quality. This is due to the fact that BAR5 was supported relatively late \citep{schneider17, debes19} since the installation of STIS on \textit{HST} in 1997, and its PSF diversity is not comparable to that of the other STIS occulters \citep[e.g., WedgeA0.6, WedgeA1.0: ][]{ren17}. Nevertheless, given that it offers the narrowest IWA in visible wavelengths \citep{debes19}, the increase of popularity in BAR5 \citep[e.g.,][]{schneider18, walker21, ren23, debes23} can help provide better PSF templates in the future. Its potential in PSF modeling can be maximized when a PSF reference does not match that of the target in complicated scenarios (e.g., source-variability-induced color and thus PSF change in STIS: \citealp{stark23}).

\section{Auxiliary images}\label{app-aux}
\subsection{Phase II planning and coverage map}
To enable \textit{HST} users in Phase II for realistic observation coverage planning with the APT for STIS coronagraphy, we have developed the {\tt VISIT-STIS-Coron} visibility tool \citep{visit-stis-coron}. In the visibility tool, the parameters are consistent with the APT conventions, including occulting location, telescope parameters (e.g., ORIENT,\footnote{The ORIENT parameter in the APT is not the \texttt{ORIENTAT} parameter in observation.}  POSTARG), etc. In addition, the occulting locations are empirically measured using the entire STIS coronagraphic archive in \citet{ren17}, which can ensure maximum telescope pointing repeatability between observation planning and execution. For STIS coronagraphic data reduction, {\tt VISIT-STIS-Coron} also includes a coronagraphic mask available in FITS format created in \citet{debes17}.

\begin{figure}[htb!]
\centering
	\includegraphics[width=0.49\textwidth]{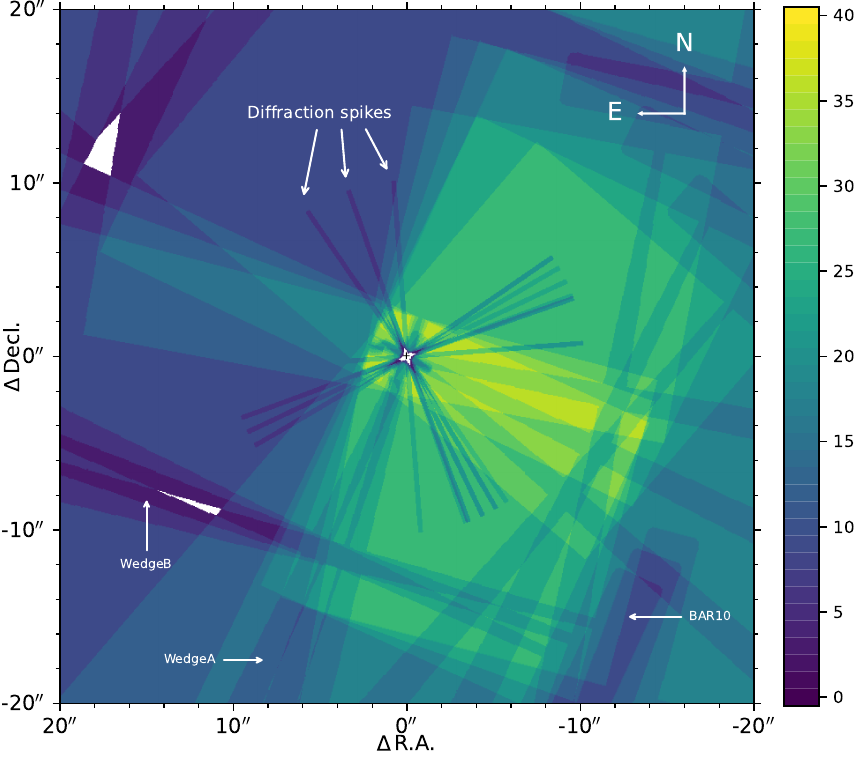}
    \caption{On-sky coverage of 3C~273 surroundings using the STIS coronagraph from 36 individual readouts in 6 orbits. The colored values denote the total readout count for a specific location. The STIS coronagraphic occulters \citep[e.g.,][]{debes19} are annotatated with arrows, and regions with zero coverage due to occulting are colored white. See Fig.~\ref{fig-3C273} for the corresponding quasar surroundings with a same field of view.}
    \label{fig-coverage}
\end{figure}

Using on-sky observations, we create the coverage map for GO-16715 in Fig.~\ref{fig-coverage} that also takes into account of the diffraction spikes of the central quasar from \textit{HST} optics. Despite a nearly $360^\circ$ azimuthal coverage, the north-east region of the quasar surroundings has less exposures, which was not originally planned in the Phase II, but instead due to an update\textsuperscript{\ref{footnote-obs}} in the \textit{HST} guide star catalog in 2022 between the two sets of 3C~273 visits.

\begin{figure}[htb!]
	\includegraphics[width=0.5\textwidth]{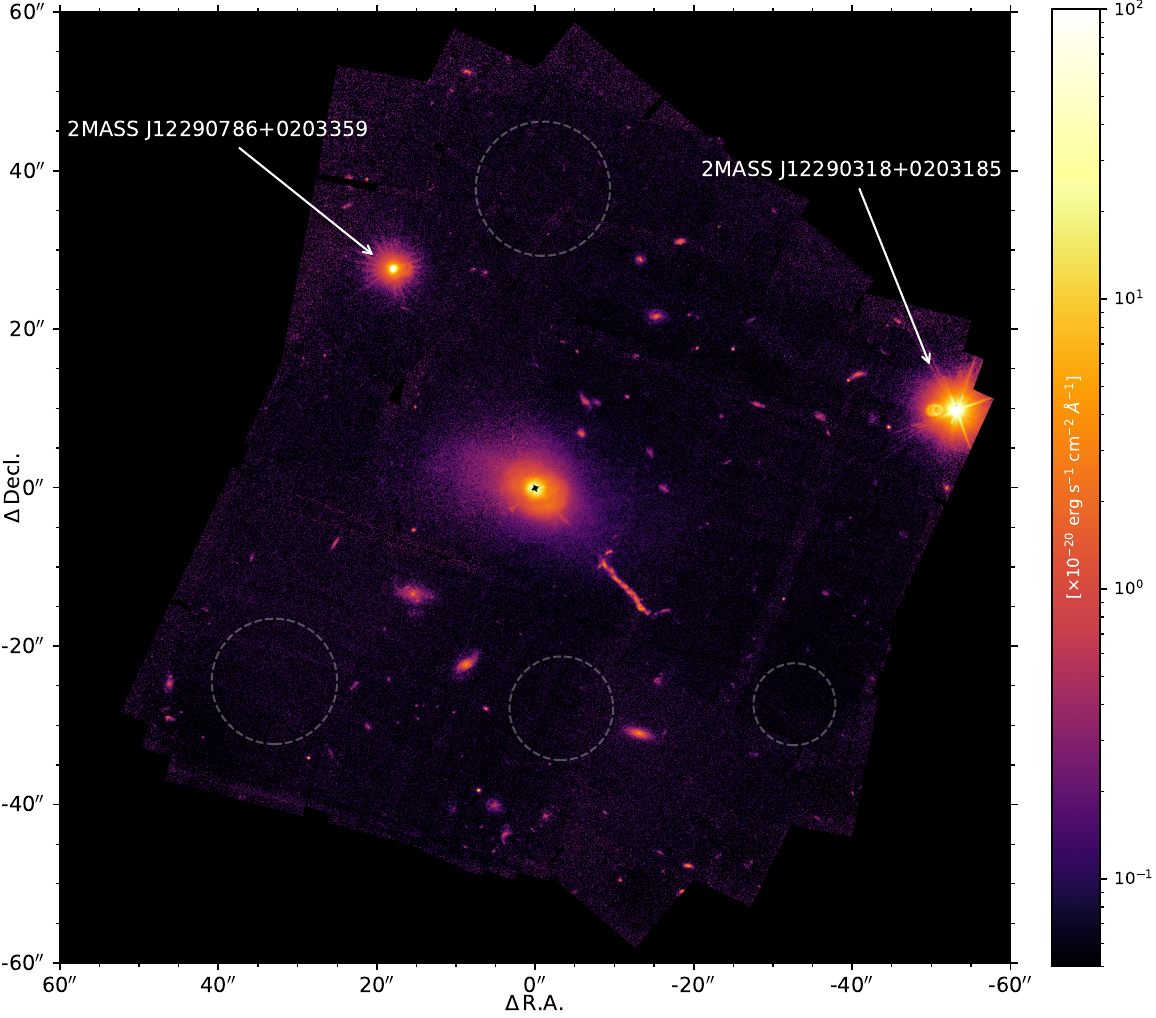}
    \caption{Full-frame result of 3C~273 surroundings from GO-16715. The regions within the four dashed circles are used to estimate the noise.}
    \label{fig-full}    
\end{figure}

\begin{figure}[htb!]
\centering
	\includegraphics[width=0.49\textwidth]{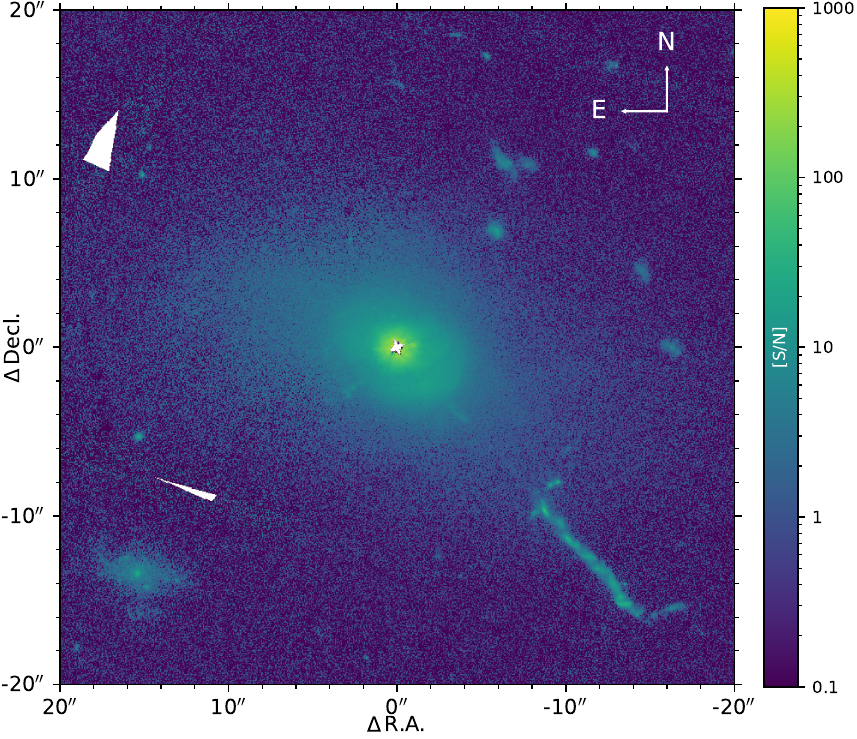}
    \caption{Pixel-wise S/N map, obtained through dividing Fig.~\ref{fig-3C273} by the pixel-wise standard deviation of the astrophysical-signal-free regions identified by eye.}
    \label{fig-sn}
\end{figure}

\subsection{Signal-to-noise map}
We calculate the pixel-wise signal-to-noise (S/N) map for Fig.~\ref{fig-3C273} as follows. First we identify four background regions which do not contain astrophysical signals by eye, see the full-frame result in Fig.~\ref{fig-full}. We then calculate the standard deviation of the selected regions to estimate the noise. The noise level does not change significantly when we used only a few regions for analysis or changed their locations and sizes. We finally divide Fig.~\ref{fig-3C273} by that calculated value of standard deviation for these background regions, and present the S/N map in Fig.~\ref{fig-sn}. 

To investigate the impact of the selection of background regions, we have changed the locations and sizes selected regions for noise estimation, and the resulting S/N maps do not have significant change from Fig.~\ref{fig-sn}. In fact, due to the positions of the BAR5 and WedgeA0.6 occulters, the telescope roll angles, and our readout in the full frame, the final full-frame result has a field of view of ${\sim}100\arcsec\times100\arcsec$. In comparison, the 3C~273 host resides within a $40\arcsec\times40\arcsec$ region in Fig.~\ref{fig-3C273} which is cropped from the full-frame result. As a result, while there are foreground stars (2MASS~J12290318+0203185 and 
2MASS~J12290786+0203359, which are located at $644\pm7$~pc and $203.4\pm1.9$~pc from the Sun in \textit{Gaia}~DR3, respectively. The two are not shown in Fig.~\ref{fig-3C273} but available in the corresponding full-frame image in Fig.~\ref{fig-full}) and background galaxies in the full-frame result, which can reach a large fraction of the areas in a $120\arcsec\times120\arcsec$ field, the majority of the full-frame result contains background regions that can be used to quantify the noise in the reduction.

\subsection{Small scale features}
\begin{figure}[htb!]
\centering
	\includegraphics[width=0.49\textwidth]{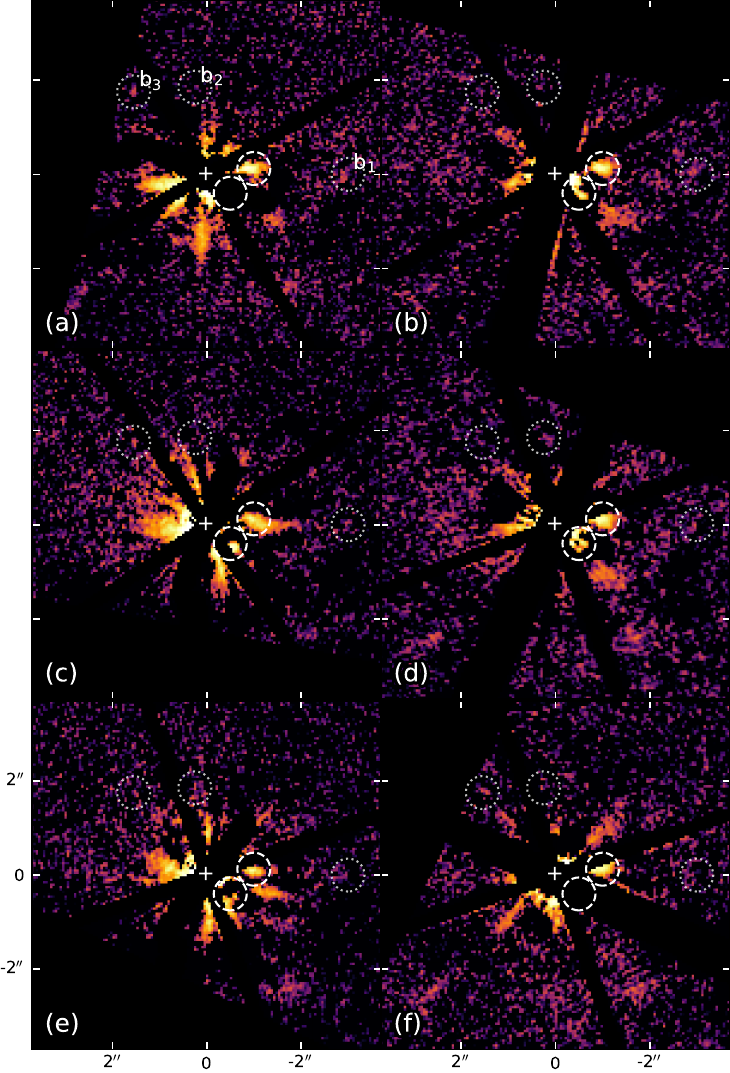}
    \caption{Small scale features in isophote-removed residuals in Fig.~\ref{fig-3C273-isophote} can persist in multiple telescope orientations. (a), (c), and (e) are for BAR5. (b), (d), and (f) are for WedgeA0.6.}
    \label{fig-features-demo}
\end{figure}

We present in Fig.~\ref{fig-features-demo} the small-scaled features identified in Fig.~\ref{fig-3C273-isophote} at different telescope orientations and occulting locations. We present the surface brightness for the newly identified features, as well as those in \citet{martel03}, from Fig.~\ref{fig-3C273-isophote} in Table~\ref{tab:sb}.

\begin{table}[htb!]
\centering
\caption{Surface brightness of structures in Fig.~\ref{fig-3C273-isophote} \label{tab:sb}}
\begin{tabular}{c|c}    \hline\hline
{Feature} & {Surface Brightness}\\
 & ($\times10^{-20}$~erg~s$^{-1}$~cm$^{-2}$~$\AA^{-1}$) \\ \hline
 b$_1$ & ${\sim}0.5$\\
  b$_2$ & ${\sim}0.3$\\
 b$_3$ & ${\sim}0.3$\\
CB & $5\pm2$ \\
CC &  $14\pm5$  \\
CJ & $3\pm2$ \\ 
filament & ${\sim}0.4$\\ \hline
E1 & ${\sim}0.6$ \\
JC & $0.37\pm0.17$ \\
IJ & $0.25\pm0.13$ \\ \hline
background & $0.013\pm0.013$\\ \hline
\end{tabular}
\end{table}

\section{Jet motion measurement}\label{app-jm}
We assume that the surface brightness distribution of an elliptical component follows a bivariate normal distribution,
\begin{equation}
S(\bm{r}_0) \sim \mathcal{N}_2\left(\bm{\mu_0}, \bm{\Sigma_0} \right),\label{eq-2d}
\end{equation}
where $\mathcal{N}_k$ denotes a normal distribution with dimension $k\in\mathbb{Z}$, $\bm{r}_0 = [x, y]^T \in \mathbb{R}^{2\times1}$ denotes the on-sky location, $\bm{\mu}_0\in \mathbb{R}^{2\times1}$ and $\bm{\Sigma}_0 \in \mathbb{R}^{2\times2}$ are the expectation and covariance matrix for the distribution, respectively. We can perform matrix translation then rotation to obtain a new surface brightness distribution $S'$.

\subsection{Elliptical component motion}

To enable the translation and rotation of the surface brightness distribution in Eq.~\eqref{eq-2d}, we define $\bm{r} = [\bm{r}_0^\top, 1]^\top$ to be a $3\times1$ column matrix, where $^\top$ denotes matrix transpose. We additionally define its corresponding expectation and covariance matrices to be 

\begin{equation}\label{eq-mu}
\bm{\mu}=\bm{\mu}_0\oplus1,
\end{equation}
and 

\begin{equation}\label{eq-Sigma}
\bm{\Sigma}=\bm{\Sigma}_0\oplus0,
\end{equation}
respectively, where $\oplus$ is matrix direct sum. With these, we can rewrite Eq.~\eqref{eq-2d} in a 3-dimensional form,

\begin{equation}
S(\bm{r}) \sim \mathcal{N}_3\left(\bm{\mu}, \bm{\Sigma} \right).\label{eq-3d}
\end{equation}

For a translation matrix $T$, we have

\begin{equation*}
T = \begin{bmatrix}
1 & 0 & t_x \\
0 & 1 & t_y \\
0 & 0 & 1 \\
\end{bmatrix},
\end{equation*}
where $t_x\in\mathbb{R}$ and $t_y\in\mathbb{R}$ denote the translation along the $x$-direction and $y$-direction, respectively. For a rotation matrix $R$, we have

\begin{equation*}
R = \begin{bmatrix}
\cos t_\theta & -\sin t_\theta & 0 \\
\sin t_\theta & \cos t_\theta & 0 \\
0 & 0 & 1 \\
\end{bmatrix},
\end{equation*}
where $t_\theta\in[-\pi,\pi)$ denotes the clockwise rotation about the origin. A translation then rotation a surface brightness distribution following Eq.~\eqref{eq-3d} then follows

\begin{equation*}
S'(\bm{r}) = R TS(\bm{r}) \sim \mathcal{N}_3\left(R T\bm{\mu}, R T\bm{\Sigma} T^\top R^\top  \right).
\end{equation*}
Given that the last row and column of Eq.~\eqref{eq-Sigma} are composed of 0, we can rewrite the above distribution as

\begin{equation}
S'(\bm{r}) \sim \mathcal{N}_3\left(R T\bm{\mu}, R \bm{\Sigma} R^\top  \right).\label{eq-3d-2}
\end{equation}

\subsection{Motion quantification}\label{app-motion-procedure}

\begin{figure}[htb!]
\centering
	\includegraphics[width=0.49\textwidth]{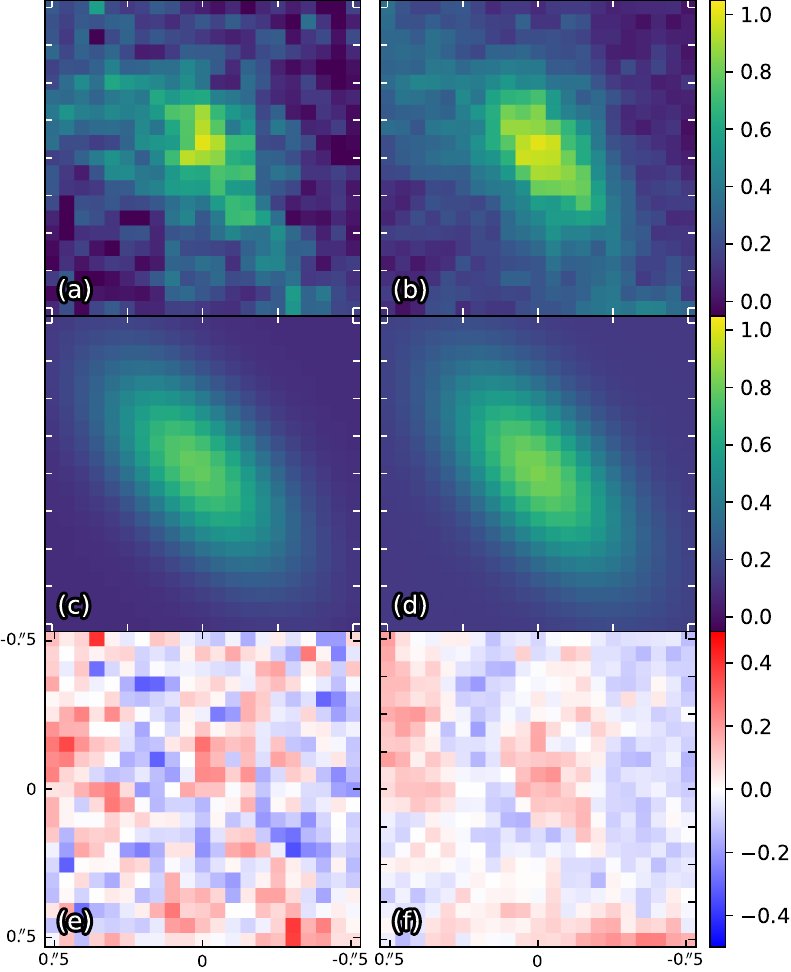}
    \caption{Ellipse fitting to obtain the morphology and motion for jet component A2. The (a) 2000 and (b) 2022 data are normalized to have a peak count of one in each panel. (c) and (d) are the best-fit models, which follow identical bivariate normal distribution but with location offset, for (a) and (b), respectively. (e) and (f) are the residuals after subtracting the models from the corresponding data.}
    \label{fig-ellipse-demo}
\end{figure}

For a bivariate normal distribution at two epochs, we assume that its morphology does not change significantly. In this way, we can use Eqs.~\eqref{eq-3d} and \eqref{eq-3d-2}, which share identical morphological parameters $\bm{\mu}$ and $\bm{\Sigma}$, to obtain the relative motion between two epochs. To maximize the information from both datasets, we use the statistical concept of dummy variables \citep[see, e.g.,][for an application in high-contrast imaging science]{ren20} to fit Eqs.~\eqref{eq-3d} and \eqref{eq-3d-2} simultaneously.

First, we rewrite the distribution of the first two entries in Eqs.~\eqref{eq-3d} in a general form in Cartesian coordinates, with the distribution centered at $(x_0, y_0)$ and rotated $\theta_0\in[-\pi, \pi)$ along the counterclockwise rotation from a rectangular bivariate normal distribution. We have

\begin{equation}\label{eq-2d-general}
S(x, y) = A \exp\left\{-\frac{1}{2}\left[a(x-x_0)^2 +2b (x-x_0)(y-y_0) + c(y - y_0)^2\right]\right\},
\end{equation}
where $A\in\mathbb{R}$ is the amplitude, and 

\begin{equation}
\left\{\begin{matrix}
a &= \frac{\cos^2\theta_0}{\sigma_x^2} + \frac{\sin^2\theta_0}{\sigma_y^2}\\
b &= - \frac{\sin2\theta_0}{2\sigma_x^2} +  \frac{\sin2\theta_0}{2\sigma_y^2}\\
c &= \frac{\sin^2\theta_0}{\sigma_x^2} + \frac{\cos^2\theta_0}{\sigma_y^2}
\end{matrix}\right..
\end{equation}

Second, we introduce a dummy variable $D\in\{0, 1\}$ to Eq.~\eqref{eq-2d-general} to denote the offset along $x$- and $y$-axis, and rotation from $\theta_0$ using the following substitution

\begin{equation}\label{eq-substitute}
\left\{\begin{matrix}
x_0 \equiv x_0 + t_x D\\
y_0 \equiv y_0 + t_y D \\
\theta_0 \equiv \theta_0 + t_\theta D
\end{matrix}\right.,
\end{equation}
to obtain the general form for the first two entries in Eq.~\eqref{eq-3d-2} when $D=1$.

Third, combining Eqs.~\eqref{eq-2d-general}--\eqref{eq-substitute}, we can obtain the general form for the first two entries in Eq.~\eqref{eq-3d} when the dummy variable $D=0$, and the two in Eq.~\eqref{eq-3d-2} when $D=1$. To obtain the morphological and offset parameters in two epochs, we now use $(x, y, D)$ as the input independent variables, and surface brightness $S(x, y, D)$ as the dependent variable. 

For each elliptical component, using the \texttt{curve\_fit} function from \texttt{scipy} \citep{scipy}, we can obtain the best-fit parameters for the elliptical morphology (i.e., $A$, $x_0$, $y_0$, $\sigma_x$, $\sigma_y$, $\theta_0$) and the motion (i.e., $t_x$, $t_y$, $t_\theta$), as well as their covariance matrix. In practice, we also had a local background brightness value that is added to Eq.~\eqref{eq-2d-general} to quantify the background difference in two epochs. We adopt the square root of the diagonal values of the covariance matrix from \texttt{curve\_fit}  as the uncertainty of the elliptical and motion parameters. We present in Fig.~\ref{fig-ellipse-demo} a demonstration of the fitting process.

\subsection{Instrument offset calibration}
\begin{figure}[htb!]
\centering
	\includegraphics[width=0.49\textwidth]{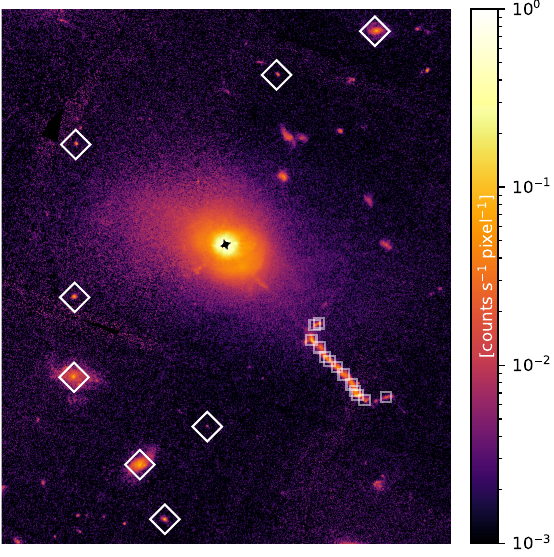}
    \caption{Components (marked by squares) and background objects (marked with diamonds) used in motion measurement and instrument offset calibration, respectively.}
    \label{fig-jet-location}
\end{figure}

To address potential bias in our alignment and rotation of two images in 2000 and 2022, we use background objects to calibrate the translation and rotation offsets between two epochs, see Fig.~\ref{fig-jet-location}. Assuming the background sources in the two images do not move, then we can first offset then rotate one image to another to calibrate the global offset.

For a background object, we can follow Appendix~\ref{app-motion-procedure} to obtain its best-fit and uncertainty values for the motion parameters: $t_x$, $t_y$, and $t_\theta$. To quantify the global offset in $\hat{t}_x$, $\hat{t}_y$, and $\hat{t}_\theta$, we assume that the motion parameters are dependent on the centers of the ellipses $(x_0, y_0)$. Using the \texttt{odr} function which performs orthogonal least-squared-fitting for data with both input and output uncertainties \citep{odr} from \texttt{scipy}, we obtained the best-fit parameters for $\hat{t}_x$, $\hat{t}_y$, and $\hat{t}_\theta$.

For a jet component, we followed Appendix~\ref{app-motion-procedure} to obtain its offset, then corrected the global offset measured from background objects. To obtain the jet component motion along the jet direction, we first obtained the position angle of the jet by performing least-square fit to the location of the jet components. We then projected the calibrated offset for a jet component, which is expressed as a bivariate normal distribution from the \texttt{odr} outputs, to the jet direction though matrix rotation, see Eq.~(B8) in \citet{shuai22} for a similar approach.

\subsection{Measured motion}

We present the measured motion rates from Sect.~\ref{sec-jet-motion} in Table~\ref{tab:motion}.

\begin{table}[htb!]
\centering
\caption{Measured motion rates along the jet in Fig.~\ref{fig-motion} \label{tab:motion}}
\begin{tabular}{c|c c c}    \hline\hline
Feature & Distance to 3C~273 & Radial Motion & Significance\\  \hline
A1    &	$13\farcs0$   &	$-3.4c\pm2.5c$   &	$1.3\sigma$ \\ 
A2    &	$14\farcs2$   &	$-0.4c\pm5.2c$   &	$0.1\sigma$ \\
B1    &	$15\farcs2$   &	$-0.6c\pm3.4c$   &	$0.2\sigma$ \\
B2    &	$15\farcs8$   &	$2.6c\pm4.1c$   &	$0.6\sigma$ \\
C1    &	$16\farcs8$   &	$-4.4c\pm5.0c$   &	$0.9\sigma$ \\
C2    &	$17\farcs8$   &	$-5.6c\pm6.1c$   &	$0.9\sigma$ \\
C3    &	$19\farcs1$   &	$7.6c\pm5.3c$   &	$1.4\sigma$ \\
D     &	$19\farcs8$   &	$2.9c\pm4.3c$   &	$0.7\sigma$ \\
H3    &	$20\farcs3$   &	$3.0c\pm4.3c$   &	$0.7\sigma$ \\
H2    &	$21\farcs2$   &	$6.0c\pm7.1c$   &	$0.8\sigma$ \\ \hline
In1\tablenotemark{a}   &	$12\farcs2$   &	$-1.5c\pm3.1c$   &	$0.5\sigma$ \\ 
In2\tablenotemark{a}   &	$12\farcs4$   &	$-6.7c\pm2.5c$   &	$2.7\sigma$ \\
Ex1\tablenotemark{a}   &	$22\farcs5$   &	$5.9c\pm6.9c$   &	$0.8\sigma$ \\ \hline
\hline
\end{tabular}
\begin{flushleft}
{\tiny \textbf{Notes}: 
$^a${Background galaxies in \citet{meyer16}.}
}
\end{flushleft}
\end{table}

\end{CJK*}
\end{document}